\documentclass[graybox]{svmult}

\usepackage{mathptmx}         % selects Times Roman as basic font
\usepackage{makeidx}          % allows index generation
\usepackage{graphicx}         % standard LaTeX graphics tool when including figure files
\usepackage{multicol}         % used for the two-column index
\usepackage[bottom]{footmisc} % places footnotes at page bottom
\usepackage{natbib}           % Natbib package for references
\usepackage{color}
\input{journalnames.sty}      % Journal names abbreviations for references

\makeindex             % used for the subject index
\begin{document}

\title*{The Stellar Kinematics of Extragalactic Bulges}
\author{Jes\'us Falc\'on-Barroso}
\institute{Jes\'us Falc\'on-Barroso \at Instituto de Astrof\'isica de Canarias,
           V\'ia L\'actea s/n, La Laguna, Tenerife, Spain\\ \email{jfalcon@iac.es}}

\maketitle

\abstract*{Galactic bulges are complex systems. Once thought to be small-scale
versions of elliptical galaxies, advances in astronomical instrumentation
(spectroscopy in particular) has revealed a wealth of photometric and kinematic
substructure in otherwise simple-looking components. This review provides an
overview of how our perspective on galactic bulges has changed over the years.
While it is mainly focused on aspects related to the dynamical state of their
stars, there will be natural connections to other properties (e.g. morphology,
stellar populations) discussed in other reviews in this volume.}

\abstract{Galactic bulges are complex systems. Once thought to be small-scale
versions of elliptical galaxies, advances in astronomical instrumentation
(spectroscopy in particular) has revealed a wealth of photometric and kinematic
substructure in otherwise simple-looking components. This review provides an
overview of how our perspective on galactic bulges has changed over the years.
While it is mainly focused on aspects related to the dynamical state of their
stars, there will be natural connections to other properties (e.g. morphology,
stellar populations) discussed in other reviews in this volume.}
\newpage

%=============================================================================
\section{Introduction}
\label{sec:1}

Galactic bulges have been generally assumed to be simple components that,
morphologically, closely resemble elliptical galaxies. First photometric
decompositions of lenticular and spiral galaxies
\citep[e.g.][]{1993MNRAS.265.1013C} established that the radial behaviour of
their surface brightness followed a de Vacouleours \citep{1948AnAp...11..247D}
or a S\'ersic profile \citep{1968adga.book.....S} with typically high $n$
values. In the mid 90s, we discovered that bulges in late-type, spiral galaxies
were smaller and displayed exponential profiles
\citep{1995MNRAS.275..874A,1996ApJ...457L..73C,1999ApJ...523..566C}. This
difference observed in the light profiles was also present in their colours,
with exponential bulges displaying bluer colours than those with larger S\'ersic
$n$ \citep[e.g.][]{2004ApJS..152..175M,2009MNRAS.395.1669G}. Despite the marked
distinction in their light profiles, the variation of colour between bulges and
their surrounding disks is rather smooth
\citep[e.g.][]{1994AJ....107..135B}.\medskip

Our view of the location of bulges in the major scaling relations (e.g.
Faber-Jackson [\citealt{1976ApJ...204..668F}], Kormendy relation
[\citealt{1977ApJ...218..333K}], or Fundamental Plane
[\citealt{1987ApJ...313...42D,1987ApJ...313...59D}]) has also evolved over time.
The sample selection biases introduced in the first studies (e.g. predominantly
early-type galaxies) showed no significant differences between bulges and
elliptical galaxies \citep[e.g.][]{1989ARA&A..27..235K, 1996MNRAS.280..167J,
2007ApJ...665.1104B}. With samples nowadays including large numbers of spiral
galaxies, our understanding of the situation of bulges in those relations has
now drastically changed \citep[e.g.][]{2009MNRAS.393.1531G,
2010MNRAS.405.1089L,2015MNRAS.446.4039E}.\medskip

\begin{figure}
\centering
\includegraphics[width=0.6\linewidth]{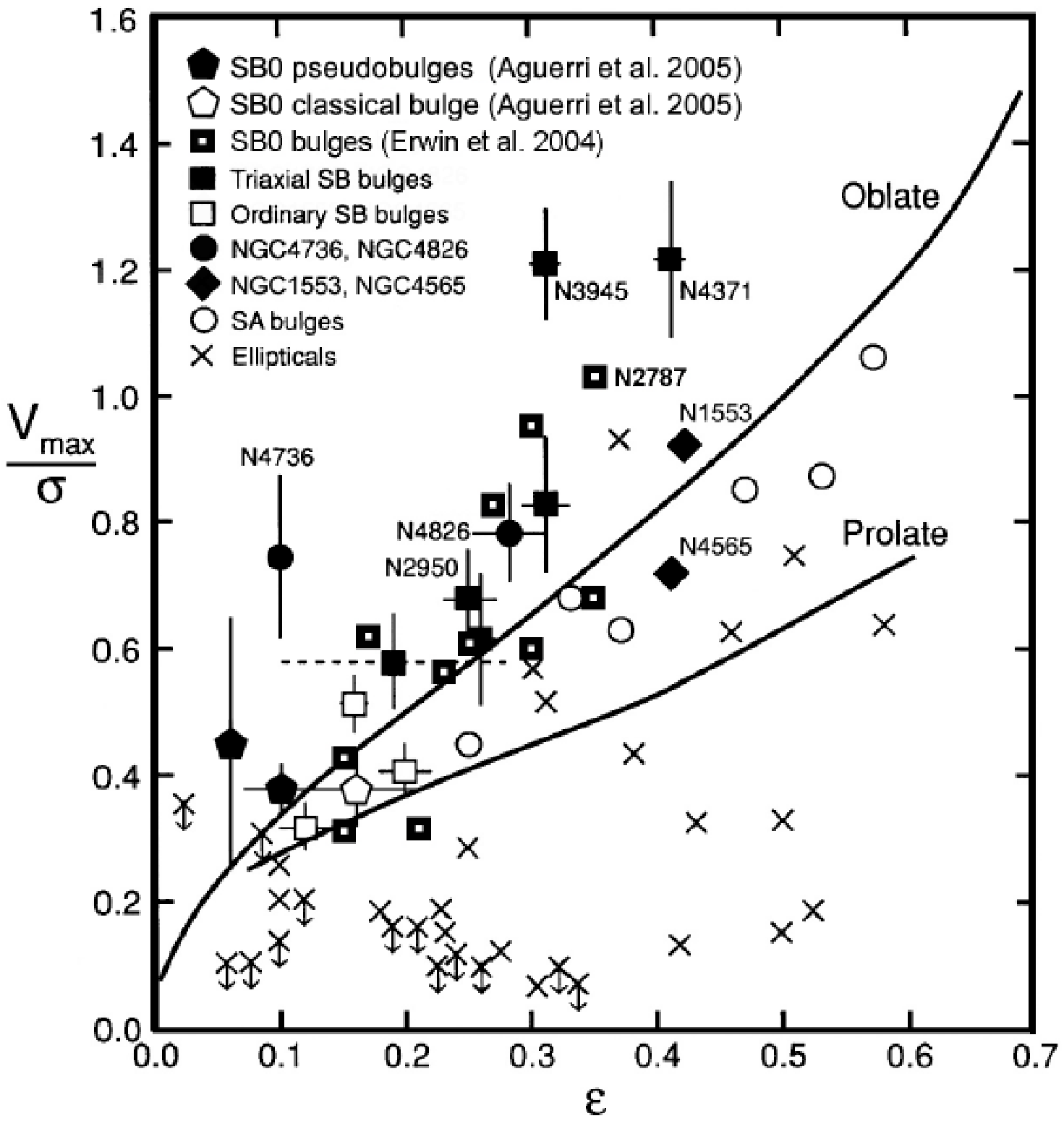}
\caption{Historical view of the level of rotational support and anisotropy of a
sample of elliptical galaxies (crosses) and bulges (remaining symbols) from
\citet{2008ASPC..396..297K}. This is an updated version of the original figure
presented in \citet{1982ApJ...256..460K}. While the physical interpretation of
this figure has evolved over time, it was the first piece of evidence suggesting
that bulges and massive early-type galaxies were intrinsically different.}
\label{fig:1}
\end{figure}

One aspect in the study of galactic bulges that has radically changed our
understanding of their nature (i.e. merger-driven structures around which disks
are formed) is their kinematics. While the photometric properties of some bulges
already pointed to a high degree of structural similarity with disks (e.g.
exponential profiles), this can only be confirmed if their kinematics also
follows that displayed by disks (e.g. significant rotation and low velocity
dispersions). In a pioneering study \citet{1982ApJ...256..460K} investigated the
degree of rotational support of a small sample of bulges compared to elliptical
galaxies. Figure~\ref{fig:1} presents an updated version, from
\citet{2008ASPC..396..297K}, of the original figure published in 1982. The
figure shows that bulges display a much larger degree of rotation than the
elliptical galaxies at a given apparent ellipticity. This was the first piece of
evidence in the literature indicating that bulges differed dynamically from
their otherwise similarly looking, slow rotating, massive early-type counterparts. While
we know now that this picture is not accurate, at the time it led to
the realisation that some bulges are actually disks and therefore may not have
formed in merger episodes, as most scenarios would assume, but rather formed
from internal material through secular processes \citep{1993IAUS..153..209K}.
These ideas evolved over time and gave rise to the definition of pseudobulges.
We refer the reader to \citet{2013seg..book.....F} for an extensive
review, produced by the lecturers of the \textit{XXIII Canary Islands Winter
School of Astrophysics}, of bulge formation and evolution in the context of
secular evolutionary processes.\medskip

In this review I will give an overview of the kinematic properties observed in
extragalactic bulges, establish their connection to the dynamical features
produced by bars, and briefly discuss the similarities with the Milky Way bulge.
I will also summarise our yet limited knowledge of the kinematics of bulges at
high redshift and end with future prospects yet to be explored in this field.

%-----------------------------------------------------------------------------
%-----------------------------------------------------------------------------

%=============================================================================
\section{Kinematic Properties of extragalactic Bulges}
\label{sec:2}

The central regions of galaxies are complex environments often displaying
multiple coexisting structural components. It is thus important to define what
we mean by a bulge in this context. In this chapter I will consider as a bulge
the stellar structures in the central regions of galaxies that ``bulge'' vertically
over the disk. The modern view is that there are three type of bulges:
classical bulges (with properties akin to elliptical galaxies), disky bulges
(with properties akin to disks), and Boxy/Peanut bulges (which are related
to bars, see \S\ref{sec:3}). In addition to bulges, the central regions of
galaxies can also host smaller structures such as nuclei, black holes, or
nuclear rings (that do not extend vertically beyond the main disk of the
galaxy).\medskip

The study of bulges is often hampered by the contamination from different
sources\footnote{It is important to remember that properties observed in
galaxies are result of integrating along the line of sight. This averaging
depends greatly on the number of components as well as the type of stars
contributing most to the light in that direction.}. In general there are two
main components that can affect our measurements: (1) the underlying main disk
of the galaxy, as so far there is no indication of truncation of disks in the
inner parts of galaxies; (2) dust, that will prevent the full integration along
the line-of-sight and thus will only allow to measure properties of stars in
front of the dust lanes. These issues are usually solved by observing galaxies
in edge-on or face-on configurations. The first one will give a clear view of
the bulge above the disk and avoid dust obscuration. It is most useful for
prominent bulges in early-type galaxies. The face-on orientation will minimize
the effects of the underlying disk. It is best for small bulges in late-type
systems, which have higher surface brightness than the disk. The drawback is
that if bulges are rotating, their signature will be likely minimal in that
orientation.\medskip

In the following subsections I will summarize the main kinematic properties
of bulges paying particular attention to those works in the literature that
have considered these issues more carefully.

\subsection{Rotational support and level of anisotropy}
\label{sec:2.1}

\citet{1982ApJ...256..460K} were the first to describe the level of rotational
support specifically in bulges of galaxies. This was achieved by measuring the
maximum rotational velocity observed in the regions above the main disk where
the light of the bulge dominates over the central velocity dispersion of the
system (V$_{\rm max}$/$\sigma$). The work by Kormendy not only concluded that
the level of rotation observed in galactic bulges was larger than that displayed
by elliptical galaxies but also, with the aid of model predictions
\citep{1981seng.proc...55B}, concluded that bulges were very likely oblate, have
isotropic velocity dispersions, and are flattened by rotation. This study was
quickly followed up by Kormendy himself \citep{1982ApJ...257...75K}, but also
other authors \citep{1983ApJ...266...41D,1983ApJ...266..516D} reaching similar
conclusions. Our current view on the level of anisotropy of bulges is, however, 
different \citep[e.g.][]{2007MNRAS.379..418C}.\looseness-2\medskip

The V$_{\rm max}$/$\sigma$--$\epsilon$ diagram has been very popular for its
power to classify dynamically different kind of galaxies, but most studies have
focused on the study of the entire systems and not in their bulge components
specifically \citep[e.g.][]{1988A&A...193L...7B,1994A&A...282L...1P,
1996ApJ...464L.119K,1999ApJ...513L..25R,2004AJ....128..121V}. With the advent of
integral field spectroscopy (IFS), this diagram has evolved and led to a
parameter (i.e. $\lambda_{\rm Re}$, \citealt{2007MNRAS.379..401E}) that allows a
more robust (and less inclination dependent) kinematic classification of
galaxies. $\lambda_{\rm Re}$ quantifies the level of specific angular momentum
in a galaxy within its half-light radius. Applied to large samples of early-type
galaxies it allowed the distinction between Slow and Fast rotating galaxies
\citep{2007MNRAS.379..401E,2011MNRAS.414..888E}. Together with model predictions
for oblate/prolate, (an)isotropic systems, it can also be used to establish the
level of anisotropy of galaxies. This aspect was explored by
\citet{2007MNRAS.379..418C} for the SAURON sample \citep{2002MNRAS.329..513D} of
early-type galaxies. This study shows that the family of Slow Rotators are
weakly triaxial, while the Fast Rotators (with V$_{\rm max}$/$\sigma$ values
similar to those observed in bulges) are typically oblate and display a wide
range of anisotropy values. The results of this study indicate that the
anisotropy observed in Fast Rotators is mainly due to a flattening of the
velocity ellipsoid in the meridional plane ($\sigma_R\ge\sigma_z$), with clear
indications that anisotropy is larger for intrinsically flatter galaxies. Given
the significant contribution of the bulge to the light in these regions, this
result suggests that bulges are actually anisotropic. This is consistent with
the level of intrinsic flattening observed in different kind of bulges  (see
M\'endez-Abreu in this volume). In this context, the study of larger samples of
bulges in late-type galaxies will be very important to fully characterize their
dynamical properties \citep[e.g. CALIFA survey,][]{2014arXiv1409.7786F}.\medskip

There has been very few attempts in the literature to extract a \textit{clean}
measurement of the anisotropy of bulges and are mostly focused on the analysis
of the Milky Way bulge. The complications to decompose accurately the
contributions of the disk to the velocity ellipsoid in the bulge dominated areas
still remains the major hurdle. The best way forward in this topic has come from
the use of detailed dynamical modelling fitting the observed stellar kinematics
\citep[e.g.][]{1991A&A...247..357B,1999A&A...349..369P,2005MNRAS.358..481K}.
Nevertheless, the main limitation of those studies is that often the shape of
the velocity ellipsoid is a property imposed in the fitting. The natural step
forward is the use of orbit-based dynamics models
\citep[e.g.][]{1979ApJ...232..236S} to separate the contributions of the bulge,
disk, and any other components present in a galaxy and thus obtain their
intrinsic properties. These models are quite demanding and require a large
number of kinematic constraints. With many IFS surveys providing data for vast
amounts of galaxies, it is only a matter of time that we exploit these analysis
tools more routinely to study the intrinsic properties of bulges.

\subsection{Scaling relations}

Many of the scaling relations used to study galaxy evolution are, in essence,
different manifestations of the Virial Theorem \citep{1870PM..40.122C}, and
relates the kinetic energy of a galaxy with the one provided by its
gravitational potential. The relationship between different
structural parameters of galaxies (e.g. absolute magnitude, half-light radius,
mean surface brightness), are discussed at length in other reviews in
this volume. Here we concentrate only on those relations that involve the
velocity dispersion of the galaxy ($\sigma$).

%-----------------------------------------------------------------------------
\begin{figure}
\centering
\includegraphics[width=0.9\linewidth]{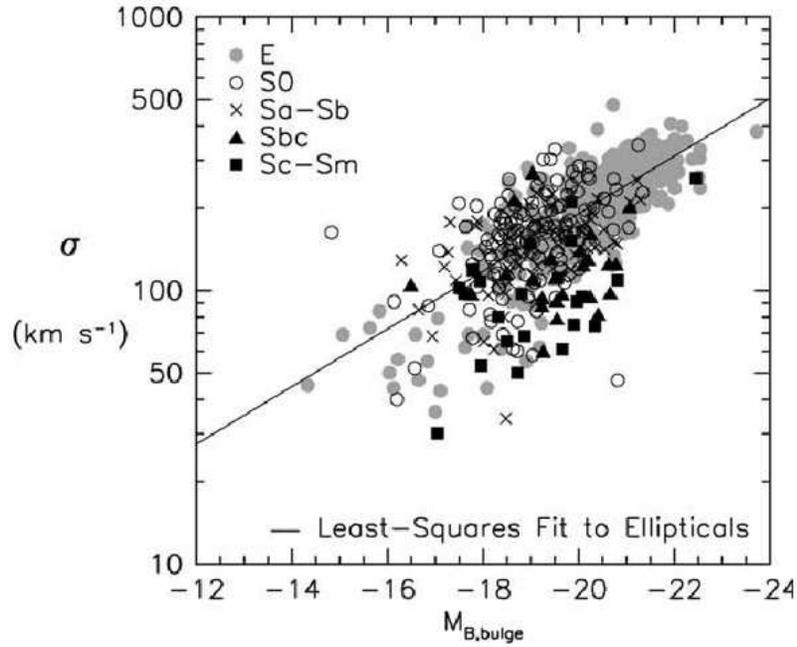}
\caption{Faber-Jackson relation for galaxies of different morphological types from
\citet{2004ASSL..319..261K}. Bulges of late-type galaxies deviate systematically
from the relation defined by ellipticals.}
\label{fig:2}
\end{figure}
%-----------------------------------------------------------------------------

\subsubsection{Faber--Jackson relation}

The Faber--Jackson relation establishes the link between the absolute magnitude
of a galaxy with its central velocity dispersion \citep{1976ApJ...204..668F}.
Early-type galaxies form a well defined sequence where more luminous galaxies
are also those exhibiting larger velocity dispersions. When it comes to the
bulges in particular, the inclusion of bulges of lenticular galaxies hardly
introduces any changes in the relation. Bulges of disk dominated spiral
galaxies, however, seem to populate different regions in this parameter space,
with largest offsets more from the relation defined by the ellipticals for those
galaxies with latest morphological types (see Figure~\ref{fig:2}). The observed
offset implies that: (1) either the bulges of later-types are brighter at a
given velocity dispersion, which would suggest the presence of younger stellar
populations (as they are also typically bluer) and/or (2) the dynamics of
late-type bulges, at a given absolute bulge luminosity, is closer to that
observed in their surrounding disks. Both cases are likely possible given that
the velocity dispersion is biased towards the younger population present along
the line-of-sight. Note, that despite the potential disky origin of those
late-type bulges, the observed relation is not driven by the luminosity of the
disk but of the bulge itself \citep[e.g.][]{2007ApJ...665.1104B}.\looseness-2

%-----------------------------------------------------------------------------
\begin{figure}
\centering
\includegraphics[angle=270,width=0.90\linewidth]{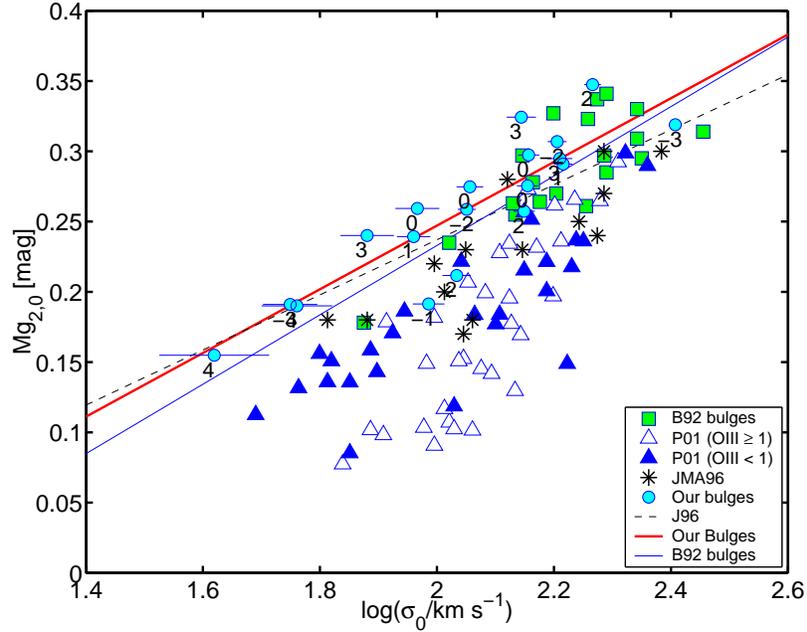}
\caption{Mg$_{2}-\sigma$ relation for galactic bulges presented in
\citet{2002MNRAS.335..741F}. The figure includes samples from this work as well
as \citet{1992ApJ...399..462B}, \citet{1996AJ....112.1415J}, and
\citet{2001A&A...366...68P}. Dashed line marks the reference relation for
early-type galaxies observed by \citet{1996MNRAS.280..167J}. Bulges of
later-type galaxies, e.g. with larger amounts of ionised-gas and younger stellar
populations, deviate most from the reference line.}
\label{fig:3}
\end{figure}
%-----------------------------------------------------------------------------

\subsubsection{Mg$_2-\sigma$ relation}

A more direct connection with stellar populations is made in the Mg$_{2}-\sigma$
relation \citep[e.g.][]{1981MNRAS.196..381T}. In Figure~\ref{fig:3} we show the
compilation made by \citet{2002MNRAS.335..741F} using their own sample together
with that of \citet{1992ApJ...399..462B}, \citet{1996AJ....112.1415J}, and
\citet{2001A&A...366...68P} against the reference relation defined for
early-type galaxies by \citet{1996MNRAS.280..167J}. Galaxies displaying larger
amounts of ionised gas (i.e. [O{\sc{iii}}] equivalent width) are also the ones
deviating most from the relation for early-types. This relation is usually
considered as a mass--metallicity relation. This is however only true in the
absence of young stellar populations. If present, the Mg$_{2}$ index is no
longer a good metallicity indicator and it becomes quite sensitive to age
\citep[e.g.][]{2010MNRAS.404.1639V}. Galaxies with large amounts of ionised-gas
are also typically the ones experiencing more intense star formation and thus
result into overall younger stellar populations. It is therefore not surprising
that the bulges in those galaxies are the ones deviating most from the relation
described by the early-type galaxies. Similar conclusions have been reached
using much larger samples \citep[e.g.][]{2002ASPC..253..321C}, although
exploring the dependence with maximum rotational velocity rather than
morphological type.

\subsubsection{Fundamental Plane relation}

The Fundamental Plane is one of the most studied scaling relations. It relates
the half-light radius of galaxies to the mean surface brightness within that
radius and the central velocity dispersion of the galaxy. As many other scaling
relations, early-type galaxies have been studied extensively
\citep[e.g.][]{1987ApJ...313...42D,1987ApJ...313...59D, 1996MNRAS.280..167J,
1998AJ....116.1591P, 1999MNRAS.304..225M, 2003AJ....125.1866B,
2008ApJ...685..875D, 2009MNRAS.396.1171H, 2010MNRAS.408.1335L,
2012MNRAS.427..245M, 2013MNRAS.432.1709C}. In contrast, the specific location of
bulges in the relation has not been explored much and has been limited to
galaxies with prominent bulges.\medskip

One of the first studies in this respect was carried out by
\citet{1992ApJ...399..462B}. They showed that bulges of lenticular galaxies
followed the relation defined by elliptical galaxies. This result was later
confirmed by \citet{2002MNRAS.335..741F}, who also found that bulges of
later-type galaxies (e.g. Sbc) were slightly displaced with respect to the main
relation. Bulges presenting the largest offsets were those with younger stellar
populations and lower velocity dispersions. These authors showed that the
offsets could be removed if one considers the missing rotational support
expected in these late-type bulges. As the rotational support of some bulges
increases, the measured velocity dispersion is no longer a reliable tracer of
their motion. In those cases rotational velocity is a much better probe of those
motions. For purely rotationally supported systems the Tully--Fisher relation
\citep{1977A&A....54..661T} is the one often the one invoked. Several studies
have confirmed that when the full kinetic energy is accounted for and
differences in the stellar populations are considered, galaxies of all
morphological types form a single relation \citep[e.g.][]{1994A&A...282L...1P,
1996A&A...309..749P, 2006MNRAS.366.1126C, 2010ApJ...717..803G,
2011MNRAS.417.1787F}, with remaining scatter typically driven by changes in
their mass-to-light ratios \citep[e.g.][]{2013MNRAS.432.1709C}.

\subsection{Radial behaviour}

The study of the kinematic radial properties of galaxies has been one of the
most prolific areas in astronomy. Mainly for bulges of early-type galaxies
\citep[e.g.][]{1982ApJ...256..460K, 1997AJ....113..950F, 1998A&AS..133..317H,
1999A&AS..136..509H, 2003A&A...405..455F, 2004MNRAS.352..721E,
2010MNRAS.408..254S}, over time we quickly started to routinely explore the
motions of stars in late-type systems \citep[e.g.][]{1989A&A...221..236B,
1992A&A...257...69B, 2001A&A...374..394V, 2004A&A...424..447P,
2005MNRAS.358..481K, 2008MNRAS.387.1099P, 2012ApJ...754...67F}. More recently,
we have started expanding our understanding of bulges through IFS (e.g. SAURON
[\citealt{2006MNRAS.367...46G}], DisKMass [\citealt{2013A&A...557A.130M}]).
While at first only rotational velocity and velocity dispersion was extracted,
the arrival of new parametrizations of the line-of-sight velocity distributions
(e.g. Gauss-Hermite expansions, \citealt{1993ApJ...407..525V}) allowed us to
identify the presence of kinematic subcomponents in galaxies (see
\S\ref{sec:2.4} for a detailed discussion). Despite displaying clear signatures
of rotational support, it is very hard to distinguish between the signal of the
bulge and underlying disk in typical rotation curves. A much more fruitful
avenue to explore is the study of the radial behaviour of the stellar velocity
dispersion. With many bulges still having a high degree pressure support (e.g.
dynamical support by random motions), it is easiest to identify the contrast
between the velocity dispersion of the disk and the bulge-dominated
regions.\medskip

\citet{1997AJ....113..950F} is one of the first studies to correlate the slope
of the observed velocity dispersion profile with general properties of their
host galaxies (e.g. central velocity dispersion, absolute magnitude, or Mg$_2$
and Fe line-strength indices). He analysed a sample of 18 lenticular galaxies
and computed the velocity dispersion gradients along the major and minor axes of
the galaxies. Compared to bright elliptical galaxies, the velocity dispersion
profiles of lenticulars in his sample were much steeper. This is expected given
that the profiles reached the low dispersion regimes observed in the disk
dominated regions. The contrast between the velocity dispersion in the bulges
and disks of his galaxies was therefore large. The intriguing result of this
study was to discover that there was no correlation between these gradients and
central velocity dispersion ($\sigma_0$), absolute magnitude or gradients of
metallicity sensitive line-strength indices. The lack of correlation with
central velocity dispersion was particularly surprising, as one would expect a
larger contrast (i.e. steeper gradient) between the very high central dispersion
galaxies and their surrounding disk. At face value, this result suggests that:
(1) the sample used in this study did not cover a sufficiently large range of
central velocity dispersion values, which could be true as the lowest $\sigma_0$
was above 100\,km\,s$^{-1}$ or (2) galaxies with dynamically hotter bulges (i.e.
with larger $\sigma_0$) have also hotter disks. At this point, with the current
sample it was not possible to discern between the two scenarios.\medskip

%-----------------------------------------------------------------------------
\begin{figure}
\centering
\includegraphics[width=\linewidth]{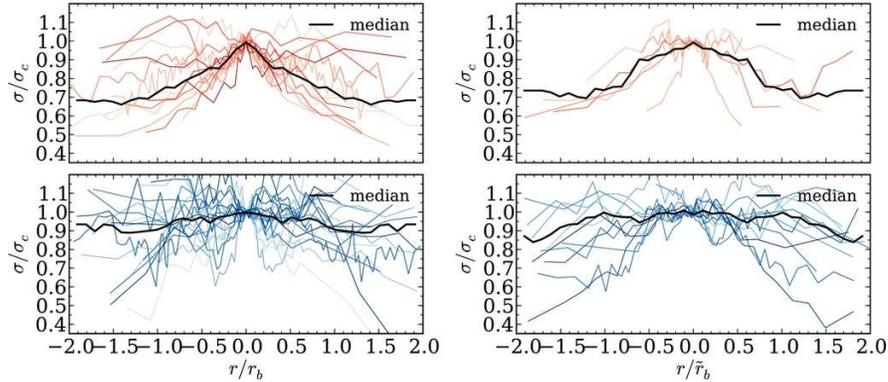}
\caption{Radial velocity dispersion profiles for a sample of 45 lenticular to
spiral galaxies from \citet{2012ApJ...754...67F}. Profiles have been normalised
to their central velocity dispersion and bulge radius. Profiles of classical
bulges are plotted in red and pseudobulges in blue. Major axis profiles are
shown on the left and minor axis on the right columns respectively. The thick
black lines correspond to the median of the individual profiles.}
\label{fig:4}
\end{figure}
%-----------------------------------------------------------------------------

The next natural step in this direction was to extend the sample to later-type
galaxies. \citet{2003A&A...405..455F} studied the radial kinematic profiles
(along the minor axis) of 19 galaxies with morphological types expanding between
S0 and Sbc. The sample was carefully chosen to have intermediate inclinations
and thus permit access to the bulge with minimal contamination of the disk on
one side of the galaxy. Central velocity dispersions ranged from 50 to over
300\,km\,s$^{-1}$. The analysis of their sample did show remarkably different
$\sigma$ radial profiles. While about half of the sample displayed very steep
profiles, the remaining set showed mainly flat profiles. The lack of velocity
dispersion gradient in a fair amount of galaxies in the sample was yet another
piece of evidence pointing to the disky nature of some galactic bulges. In
relation to the properties of the host galaxy, there was a slight tendency for
galaxies with flatter profiles to display higher disk central surface
brightness. A trend was also found with the ellipticity of the bulge component
in the sense that more flattened bulges showed shallower gradients. Despite
analysing galaxies covering a wider range of morphological types, no correlation
was found with either morphological type index, bulge S\'ersic index $n$, bulge
and disk scale lengths and bulge effective surface brightness. It appears that
the disky nature of bulges cannot be established on the basis of spheroid
luminosity, as velocity dispersion gradients do not seem to correlate with
bulge luminosity or with central velocity dispersion either.\looseness-2\medskip

\citet{2012ApJ...754...67F} presents the most recent effort in the literature
trying to address these issues. In this work 45 S0 to Sbc galaxies were studied
with the goal of relating the kinematic information with photometric properties
typical of classical and pseudobulges\footnote{Note that in this work the
definition of a bulge differs from the one used in this review. While
\citet{2012ApJ...754...67F} define bulges as structures with flux above
the disk surface brightness profile, here they are also required to extend
vertically above the disk.}. The sample contained a fair fraction of barred
galaxies and displayed a wide range of central velocity dispersions (between
$\sim$50 to 200\,km\,s$^{-1}$) and absolute magnitudes (from $-18$ to
$-21$\,mag). The galaxies were also moderately inclined with allowed access to
the bulge region without being significantly affected by dust in the disk.
Figure~\ref{fig:4} shows the radial behaviour of the velocity dispersion along
the major and minor axes of the galaxies in the sample. Similarly to
\citet{2003A&A...405..455F}, bulges exhibit two types of profiles: steep and
flat velocity dispersion profiles. This work provides first tentative evidence
for a correlation between the slope of the velocity dispersion profile and the
bulge's S\'ersic index $n$.\medskip

The study of the stellar kinematics of late-type galaxies has usually been
hampered by complex, often dusty, morphologies. Furthermore, bulges in those
galaxies are not particularly bright which makes the extraction of any
spectroscopic measurement (kinematic in particular) specially harder. With the
advent of integral-field spectroscopy, a few studies have allowed a kinematic
characterisation of bulges in galaxies from Sb to Sd types.
\citet{2006MNRAS.367...46G} carried out SAURON observations of 18 spiral
galaxies with good \textit{Hubble Space Telescope} photometry available. The
velocity dispersion profiles of the galaxies were mostly flat or with positive
gradients. Very few galaxies displayed negative gradients. When looking for
correlations between these gradients and the morphological type of the galaxies,
there was only a slight tendency for earlier types to displayed negative
gradients. Positive gradients were not strongly correlated with latest Hubble
types.\medskip

The study of velocity dispersion gradients will be soon expanding thanks to the
large number of IFU surveys (DiskMass, \citealt{2010ApJ...716..198B}; CALIFA,
\citealt{2012A&A...538A...8S}]; SAMI, \citealt{2012MNRAS.421..872C}; MaNGA,
\citealt{2015ApJ...798....7B}). However, it is important to remember that not
all of them will allow the study of bulges in late-type galaxies due to
restrictions in the spatial sampling or their spectral resolution.

\subsection{Amount of substructure}
\label{sec:2.4}

So far in this review we have exposed the properties of different kind of
bulges, and yet this has gone as far as showing that some bulges exhibit
kinematics closer to what it is observed in a disk (e.g. rotation dominated)
instead of the classical idea of bulges being pressure supported. Here we will
revise the kinematic properties of the different structural components
dominating the light in the inner regions of galaxies.\medskip

Counter-rotating components are common in galaxies. Large, kpc-scale,
kinematically decoupled components (KDCs) are typically found in bright
elliptical galaxies \citep[e.g.][]{1988A&A...202L...5B, 1989ApJ...344..613F,
1997ApJ...481..710C, 1999MNRAS.306..437H,
2001ApJ...548L..33D,2014MNRAS.445L..79E}. They usually contain old stellar
populations and are almost indistinguishable from the remaining body of the
galaxy. Smaller decoupled components are, however, harder to identify, are made
of young stars and reside in lower luminosity early-type galaxies
\citep[e.g.][]{2006MNRAS.373..906M}. Large-scale counter-rotation of disk
components seems also not so rare: NGC\,4550 \citep[e.g.][]{1992ApJ...394L...9R,
1992ApJ...400L...5R}, NGC\,4138 \citep{1996AJ....112..438J}, NGC\,4473
\citep{2004cbhg.sympE...5C}. See \citet{2011MNRAS.414.2923K} for other cases
detected through a \textit{kinemetry} analysis \citep{2006MNRAS.366..787K}. The
detection of such extreme cases keeps increasing as new kinematic decomposition
techniques are developed \citep[e.g.][]{2013A&A...549A...3C,
2013MNRAS.428.1296J, 2014A&A...570A..79P}.\medskip

Counter-rotation of bulges is an odd phenomenon. There are very few cases
reported in the literature of bulges rotating around a completely different axis
than their surrounding disks. One of those striking cases is NGC\,4698
\citep{1999ApJ...519L.127B}, where the bulge appear to rotate perpendicular to
the stellar disk. Another unusual case is that of NGC\,7331, where the bulge was
reported to counter-rotate with respect to the disk
(\citealt{1996ApJ...463L...9P}, but see \citealt{1999A&A...348...77B}).
Numerical simulations suggest mergers of galaxies as the only viable path for
the formation of such structures \citep[e.g.][]{1998ApJ...505L.109B,
1998ApJ...506...93T}.\medskip

%-----------------------------------------------------------------------------
\begin{figure}
\centering
\includegraphics[width=\linewidth]{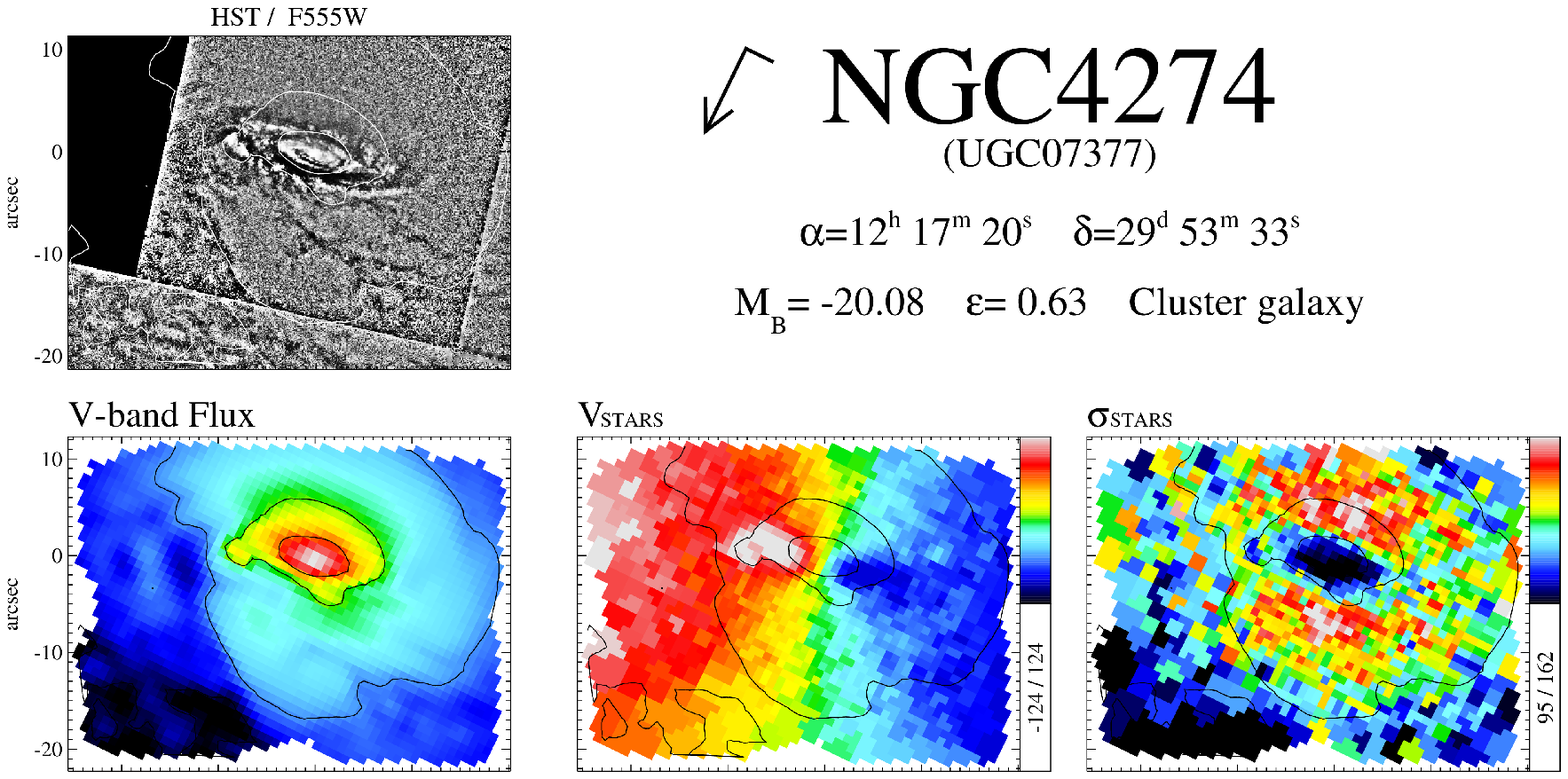}
\includegraphics[width=\linewidth]{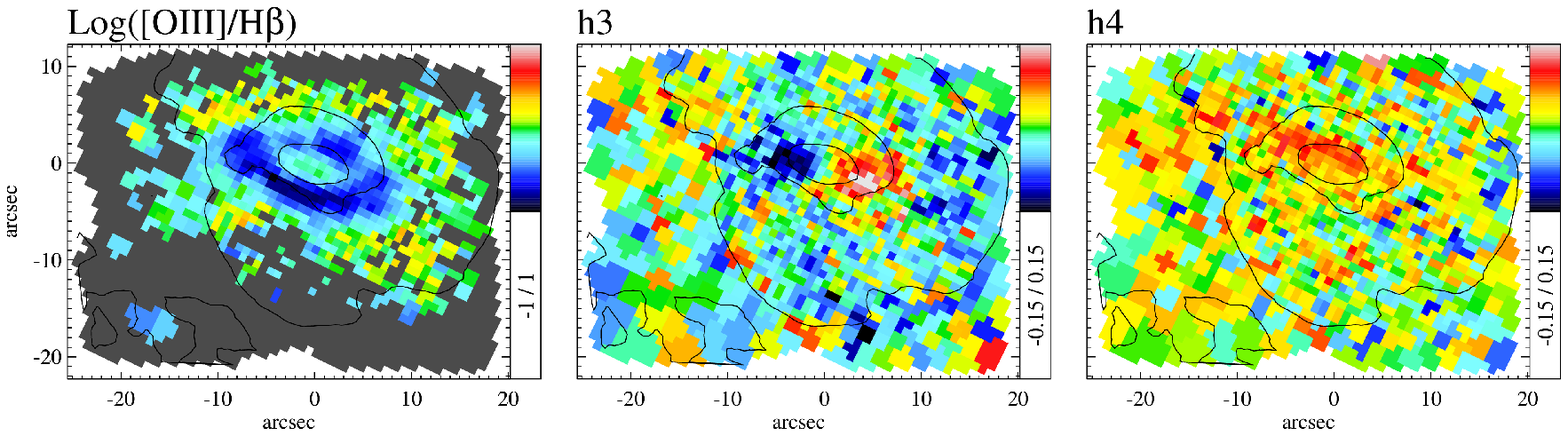}
\caption{Stellar kinematic maps for NGC\,4274 from \citet{2006MNRAS.369..529F}.
The arrow and its associated dash at the top of each figure mark the north and
east directions, respectively. (\textit{First row}) HST unsharp-masked image of
the galaxy and some basic information. (\textit{Second row}) reconstructed total
intensity (in mag/arcsec2 with an arbitrary zero point), stellar mean velocity
V, and stellar velocity dispersion in km\,s$^{−1}$). (\textit{Third row})
[O{\sc{iii}}]/H$\beta$ emission line ratio map (in logarithmic scale), and
Gauss–-Hermite moments h3 and h4 of the stellar line-of-sight velocity
distribution.}
\label{fig:5}
\end{figure}
%-----------------------------------------------------------------------------

A common feature is the presence of co-rotating components (e.g. a nuclear disk)
embedded in an otherwise pressure supported spheroidal bulge . The key kinematic
signature of these inner disks is a steep rise of the rotation velocity in the
inner parts (i.e. faster than the expected rise of the main disk) accompanied by
low velocity dispersions values. There is often also an anti-correlation between
the velocity and h$_3$ moment in the locations with lowest velocity dispersion,
which is usually an indication of multiple kinematic components. All these
features are shown in Figure~\ref{fig:5} using the two-dimensional kinematic
maps of NGC\,4274 from \citet{2006MNRAS.369..529F} as an example. The
\textit{Hubble Space Telescope} unsharped-masked image reveals the presence of a
dusty disk in the inner regions of the galaxy, which is not so obvious in the
reconstructed image of the galaxy. The disk has a clear signature in the
velocity map, and even more so in the velocity dispersion which is much lower
than the values of the surrounding dynamically hot bulge. In this particular
case, the very low [O{\sc{iii}}]/H$\beta$ emission line ratio suggests star
formation is taking place in the inner disk. The presence of these co-rotating
components do not always imply associated young stellar populations. The stellar
population analysis carried out by \citet{2007MNRAS.379..445P} of the
\citet{2006MNRAS.369..529F} sample of 24 Sa galaxies concluded that about half
of the galaxies displaying low central velocity dispersion values (so called
$\sigma$-drops, \citealt{2001A&A...368...52E,2003A&A...409..469W}) have mean
luminosity weighted ages above 5\,Gyr. The incidence of $\sigma$-drops in this
sample was about 50\%. $\sigma$-drops are not only produced by nuclear disks,
but can also be caused by nuclear dust spirals and star-forming rings
\citep{2008A&A...485..695C}. The origin of these components is often related to
the inflow of gas, driven by bars, towards the inner regions of galaxies
\citep[e.g.][]{2005MNRAS.358.1477A}. Note, however, that minor mergers could be
also responsible for the formation of inner disks and rings in spiral galaxies
\citep[e.g.][]{2011A&A...533A.104E}.

%=============================================================================
\section{Relating Bars and Bulges}
\label{sec:3}

Bars are prominent components of galaxies, produced by disk instabilities, that
can pump disk material above the plane generating central structures that also
{\it bulge} over the thin disk \citep[e.g.][]{1993ApJ...409...91H}. As we
discuss in this section, the kinematic properties of these bars are different
from those observed in common bulges. The origin of some type of bulges (e.g.
pseudobulges) appears to be tightly connected to secular evolutionary processes
induced by bars \citep[see][for a theoretical view of bulge formation in the
context of bars]{2005MNRAS.358.1477A}. Bars are active agents in the inflow of
gas towards the inner regions of galaxies \citep[e.g.][]{1999ApJ...525..691S}.
This naturally allows the formation of new structures (e.g. bulges, rings, inner
disks, central mass concentration).\medskip

\begin{figure}
\centering
\includegraphics[width=0.7\linewidth]{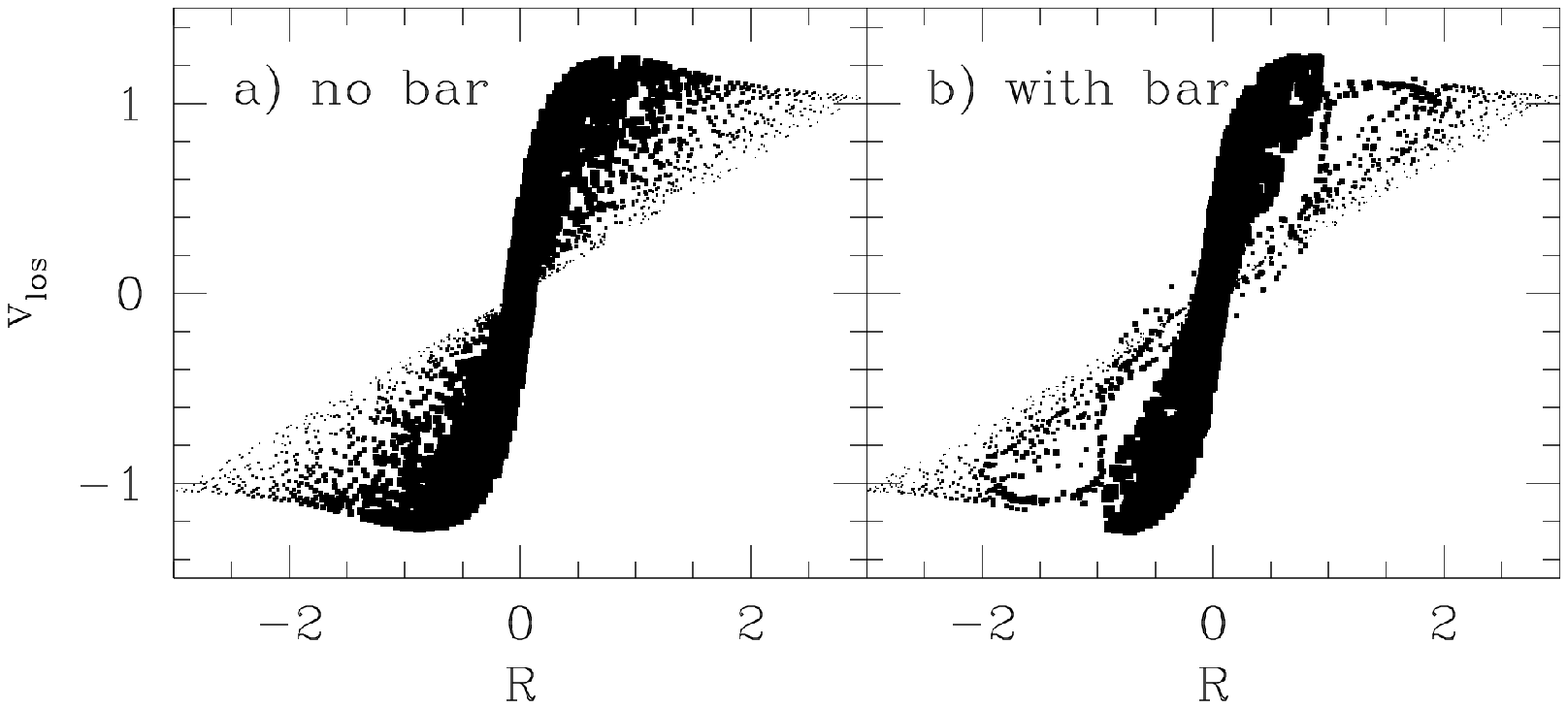}
\includegraphics[width=0.7\linewidth]{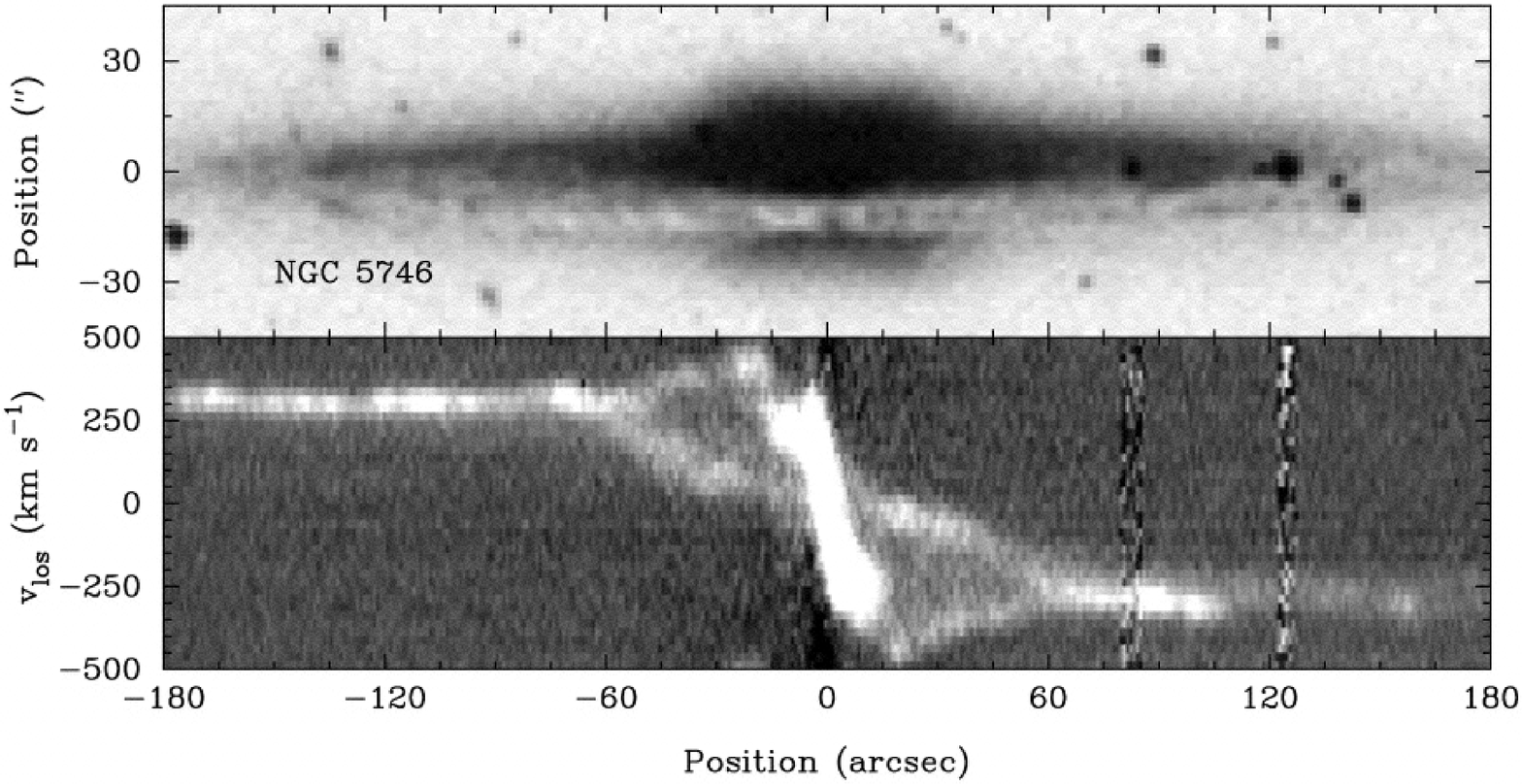}
\caption{Position--Velocity diagrams (PVDs) of barred galaxies. (\textit{Top})
Model prediction for the observed line-of-sight velocity distribution as a
function of radius for non-barred and barred galaxies
\citep{1995ApJ...443L..13K}. (\textit{Bottom}) Observed PVD for the boxy/peanut
bulge of NGC\,5746 \citep{1999AJ....118..126B}. The kinematic signature of a bar
in the observations is very evident.}
\label{fig:6}
\end{figure}

The vertical extent of bars is best observed in edge-on galaxies. When the long
axis of the bar is perpendicular to our line-of-sight bars are usually called
Boxy/Peanut (BP) bulges due to their peculiar shape. Most of the material
outside the disk plane has been elevated through bar buckling episodes early in
the evolution of the bar \citep[e.g.][]{2006ApJ...637..214M}. Kinematically, BP
bulges produce a characteristic signature (i.e. a ``figure-of-eight'') in the
Position--Velocity Diagram (PVD). This was first predicted by
\citet{1995ApJ...443L..13K} (see Figure~\ref{fig:6}, top row). With the aid of
analytical models, they determine the location of particles in this diagram for
barred and non-barred galaxies. In their view, the gap observed in the PVD of
barred galaxies is produced for a lack of available orbits near the corotation
radius of the bar. This effect should affect both the stellar and gas components
of galaxies. This prediction was nicely confirmed with larger samples of
galaxies \citep[e.g.][]{1999A&A...345L..47M, 1999AJ....118..126B}. In the case of
\citet{1999AJ....118..126B}, they produced PVDs for a sample of 30 edge-on
spiral galaxies with prominent BP bulges. Figure~\ref{fig:6}, bottom row, shows
the observed PVD for NGC\,5746 that clearly displays the predicted gap.\medskip

%-----------------------------------------------------------------------------
%-----------------------------------------------------------------------------

%-----------------------------------------------------------------------------
\begin{figure}
\centering
\includegraphics[width=0.45\linewidth]{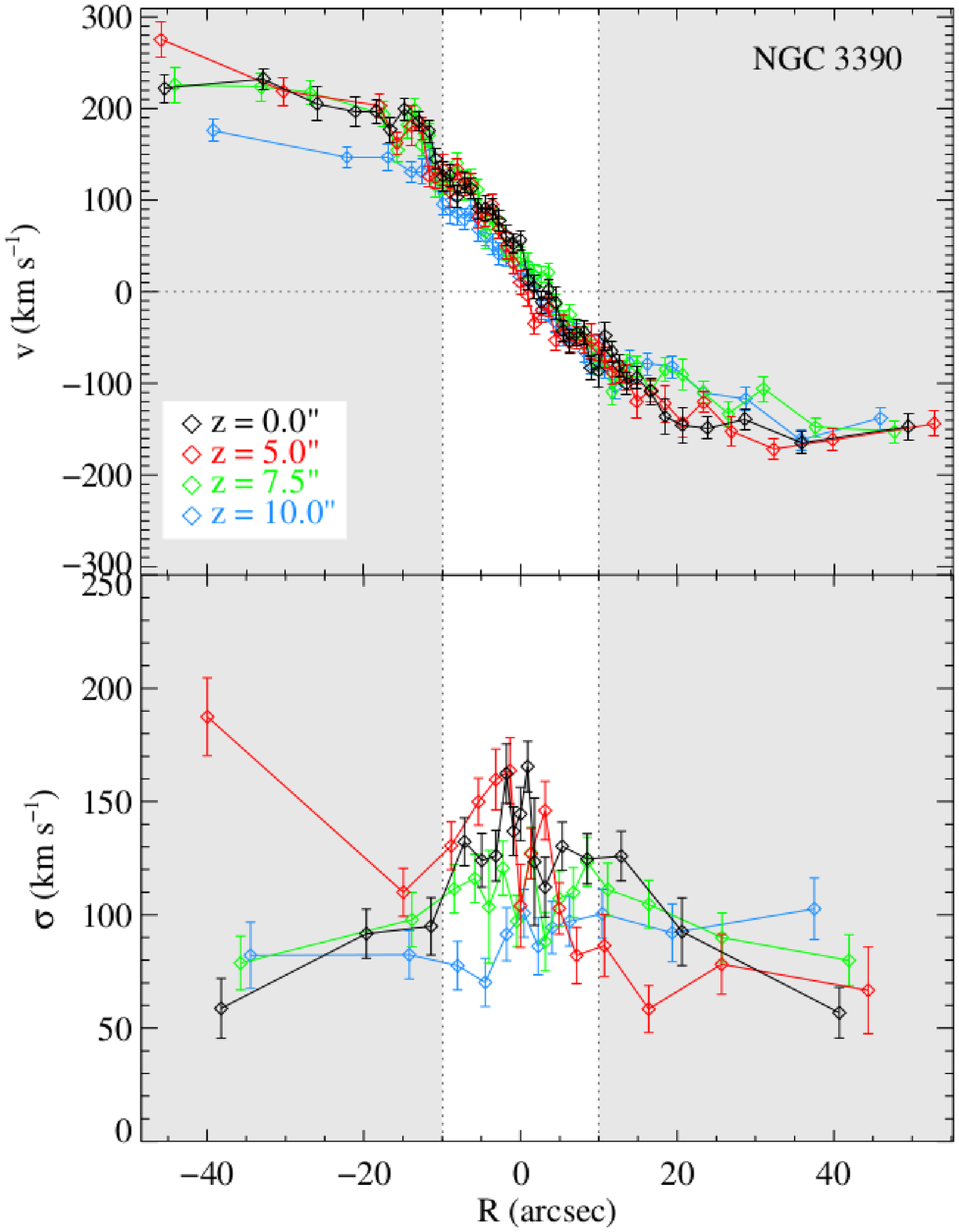}
\includegraphics[width=0.45\linewidth]{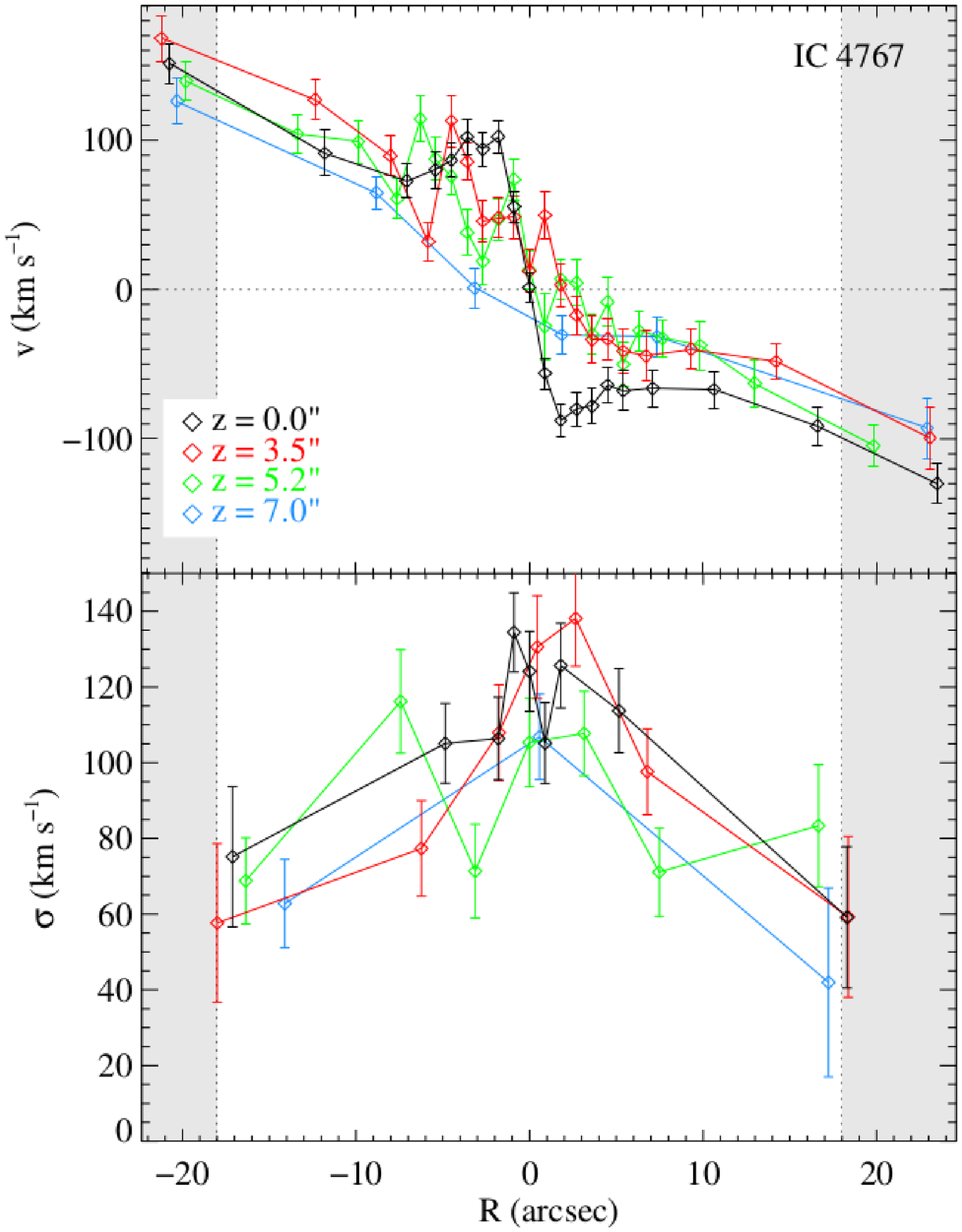}
\caption{Stellar line-of-sight rotation curves and velocity dispersion profiles
for two Boxy/Peanut, edge-on galaxies in the \citet{2011MNRAS.414.2163W} sample.
NGC\,3390 shows clear signatures of cylindrical rotation, while IC\,4767 does
not (i.e. kinematics at increasing distance from the main disk shows different
behaviour). The shaded regions mark the disk dominated regions.}
\label{fig:7}
\end{figure}
%-----------------------------------------------------------------------------

Another typical kinematic feature of BP bulges predicted by numerical
simulations is cylindrical rotation \citep[e.g.][]{1988ApJ...331..124R,
1990A&A...233...82C}. The first evidence for cylindrical rotation in galaxies
was revealed by \citep{1982ApJ...256..460K} for NGC\,4565 when studying the
stellar kinematics of galactic bulges. References of cylindrical rotation in
other galaxies are rather scarce in the literature:  IC\,3370
\citep{1987AJ.....94...30J}, NGC\,1055 \citep{1993A&A...280...33S}, NGC\,3079
\citep{1993A&A...268..511S}, NGC\,5266 \citep{1987ApJ...313...69V}, NGC\,7332
\citep{1994AJ....107..160F}. This lack of cases is likely due to: (1)
inclinations effects. Cylindrical rotation is best observed in edge-on galaxies
\citep[e.g][]{2002MNRAS.330...35A}, (2) the fact that most observations with
long-slit spectrographs targeted the major and/or minor axes of the galaxies,
which makes it difficult to detect. The most recent work addressing this aspect
of BP bulges is that of \citet{2011MNRAS.414.2163W}. This study placed long
slits parallel to the major axis of five known BP bulges. The surprising result
of this study is that not all BP bulges displayed cylindrical rotation.
Figure~\ref{fig:7} shows the analysis for two distinct cases in their sample.
While NGC\,3390 displays clear signatures of solid-body rotation, IC\,4767 presents
shallower major axis velocity profiles as a we move away from the disk. This
outcome requires further confirmation using larger samples of edge-on
galaxies. It will also benefit from studies making use of integral-field
spectrographs to map the full two-dimensional kinematics over the BP dominated
region. A glimpse of what this kind of studies can bring is
presented in \citet{2004MNRAS.350...35F} for the known case of
NGC\,7332.\medskip

\begin{figure}
\centering
\includegraphics[width=0.98\linewidth]{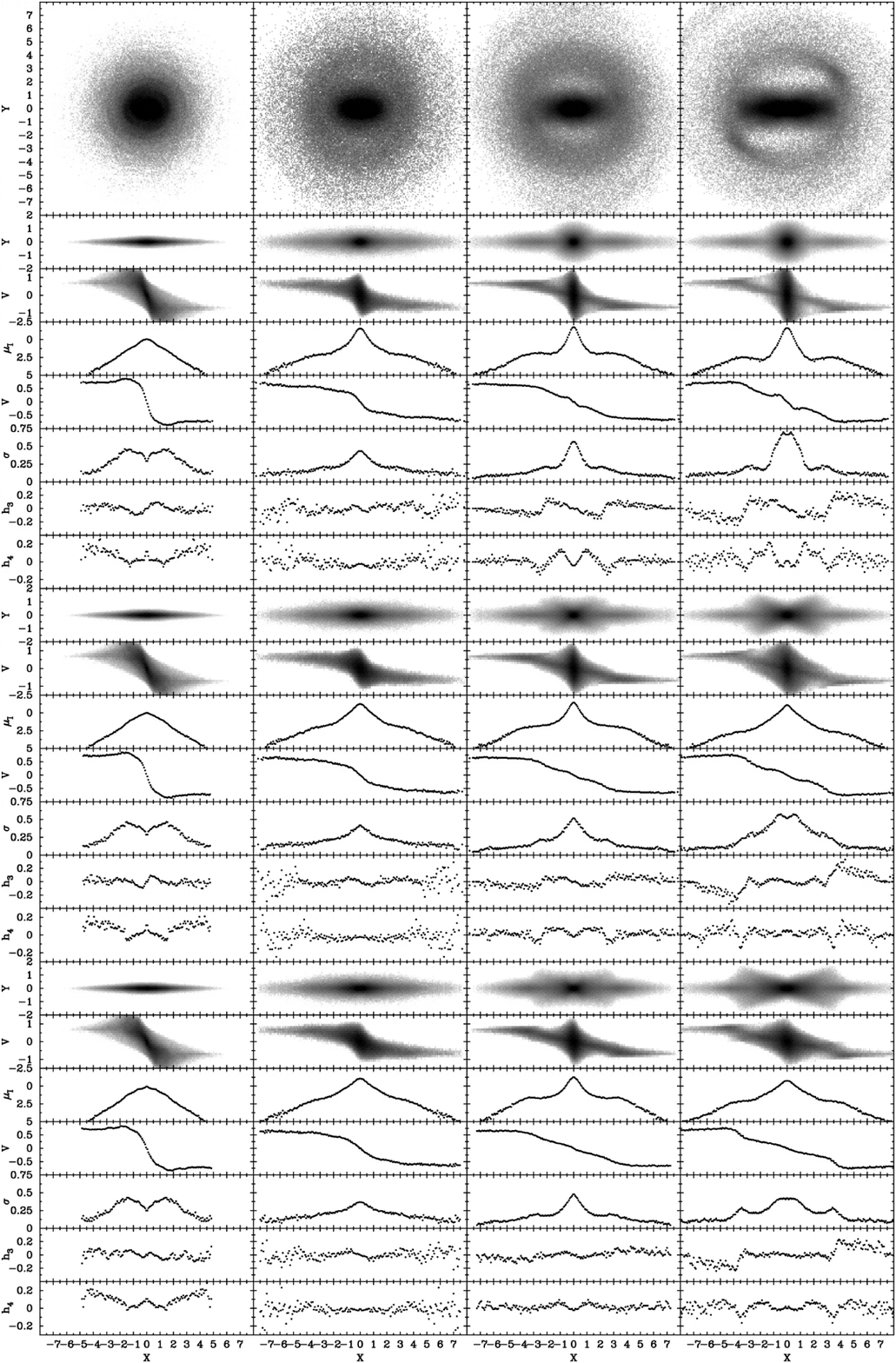}
\caption{Stellar kinematic diagnostics for barred galaxies in N-body simulations
from \citep{2005ApJ...626..159B}. (\textit{Left to right}) No-bar, weak-bar,
intermediate-bar, and strong-bar case. (\textit{Top to bottom}) image, PVD,
surface brightness, and kinematic parameters (velocity, velocity dispersion,
h$_3$ and h$_4$ Gauss-Hermite moments) as a function of bar orientation, from
end-on to side-on.}
\label{fig:8}
\end{figure}

Bars are also capable of producing other distinct features in the stellar
kinematics of galaxies, which are often related to resonances induced by the bar
itself in the host galaxy. \citet{2005ApJ...626..159B} established, using N-body
simulations, a series of kinematic diagnostics for bars of different strength
and orientations in highly-inclined galaxies (see Figure~\ref{fig:8}): (1) a
``double-hump'' rotation curves, (2) velocity dispersion profiles with a plateau
at moderate radii, and often displaying a $\sigma$-drop in the centre, (3) a
positive correlation between the velocity and the h$_3$ Gauss-Hermite moment
over the length of the bar. Some of these features have been recognised
observationally in several studies \citep[e.g.][]{1981ApJ...247..473P,
1983ApJ...275..529K, 1997A&AS..124...61B, 2001A&A...368...52E,
2003A&A...409..459M, 2009A&A...495..775P}. While having the most potential to
unravel the presence of bars, the V--h$_3$ correlation has been hardly studied
observationally \citep[e.g.][]{2004AJ....127.3192C}. These diagnostics work best
for edge-on galaxies. The kinematic tracer of BP bulges in face-on systems is
the h$_4$ Gauss-Hermite moment. Simulations carried out by
\citet{2005ApJ...628..678D} predict that a negative double minima around the
centre of the galaxy is an excellent indicator of a BP bulge for a wide range of
bar strengths and inclinations. Although the observational requirements to
measure this parameter are very demanding, this feature has been nicely
confirmed observationally by \citet{2008ApJ...679L..73M}. Interestingly,
\citet{2014MNRAS.444L..80L} suggest that the barlenses observed in the face-on
view of many disk galaxies \citep[e.g.][]{2011MNRAS.418.1452L} are effectively
the thick part of the BP bulge when seen face-on. See also
\citet{2014arXiv1405.6726A} for a theoretical interpretation.\medskip

%-----------------------------------------------------------------------------
%-----------------------------------------------------------------------------

There are strong indications that large bulges can have an effect in the
strength of a bar. Stronger bars appear in galaxies with low bulge-to-total
ratios and central velocity dispersions \citep{2008Ap&SS.317..163D,
2009A&A...495..491A, 2009ApJ...692L..34L}. What it is not well established yet,
observationally, is the effect a bar would have on the dynamics of a
pre-existing bulge. Numerical simulations by \cite{2013MNRAS.430.2039S} suggest
that a pressure supported bulge would gain net rotation as a result of angular
momentum exchange with the bar. Rotation of the final composite classical and BP
bulge would be close to cylindrical, with small deviations in the early phases
of the secular evolution. Therefore, untangling the intrinsic properties of
bulges in barred galaxies is a very difficult task that will require detailed
dynamical modelling of high quality observations. Numerical tools like the
NMAGIC code \citep{2007MNRAS.376...71D} applied to high-quality, integral-field
data \citep[e.g.][]{2013MNRAS.429.2974D} seems the way forward.\medskip

The Milky Way bulge is the most vivid example of a complex system. Besides
cylindrical rotation, it displays many of the other kinematic signatures of bars
summarised above. The origin of the multiple substructures present at the centre
of our Galaxy (possibly including other types of bulges, e.g.
\citealt{2014ApJ...787L..19N}) cannot be solved by inspecting the kinematics
alone, as angular momentum transfer is expected between them. Most of the
efforts today to solve this puzzle come from relating the observed kinematics to
the distinct stellar populations present in those regions. We refer the reader
to Oscar Gonz\'alez and Dimitri Gadotti's review in this volume for a
comprehensive summary of the properties observed in the Galactic bulge, but also
Juntai Shen's chapter for a theoretical view on the possible paths for its
formation and evolution.

%=============================================================================
\section{Kinematics of Bulges at High Redshift}
\label{sec:4}

With typical sizes of a few kiloparsecs, bulges in nearby galaxies would be very
difficult to resolve spatially at intermediate to high redshifts even with the
best instruments on board of \textit{Hubble Space Telescope}. In addition, the
morphologies of galaxies are known to deviate from the standard Hubble sequence
from redshift $\sim$1 onwards \citep[e.g.][]{2008ApJ...688...67E}, so we should
probably not think of bulges at high-redshift in the same way we think of them
in the local Universe. Nevertheless knowing the conditions, in terms of
rotational support, of the galaxies that will eventually lead to lenticular and
spiral galaxies nearby, can help us understand the kind of progenitors that will
host the variety of bulges we see today.\medskip

In the light of the large amount of pseudobulges observed in the nearby
Universe, a logical question to ask is: do we see the signatures of secular
evolution in bulges at high-$z$? Numerical simulations reproducing the clumpy
galaxies from redshift $z$\,$\sim$\,1 suggest that bulge kinematics is not very
different from the values observed for pressure-supported systems, with
(V/$\sigma$) values below 0.5 \citep[e.g.][]{2007ApJ...670..237B,
2008ApJ...688...67E}. This is likely due to the turbulent nature of clumps
merging at the centre of galaxies \citep[e.g.][]{2012MNRAS.420.3490C}. Note,
however, that the merging and migration of clumps towards the inner regions is
an internal process, as it takes place in the disk of galaxies. The physical
conditions, in terms of gas supply, for bulge formation at high redshifts are
very different from the ones observed in the local Universe. Secular evolution
takes place at a much faster pace at high-$z$.\medskip

Integral-field observations of galaxies at increasing redshifts confirm the
turbulent nature of disks, as revealed by the systematically high velocity
dispersion values \citep[e.g.][]{2013ApJ...767..104N, 2014arXiv1409.6791W}.
Nevertheless, galaxies show a wide range of kinematic properties: from well
behave rotating disks, to dispersion dominated systems, and galaxies with
chaotic motions \citep[e.g.][]{2008A&A...477..789Y, 2008ApJ...687...59G,
2011MNRAS.417.2601W, 2014MNRAS.439.1494B}. Recent results from the KMOS3D survey
\citep{2014arXiv1409.6791W} show that most galaxies, in the main star forming
sequence, between redshifts 1 and 2 are rotationally-supported. When combined
with other datasets, they measure an evolution of the ionised-gas velocity
dispersion which is consistent with the observed changes in the gas fractions
and specific star formation rates of galaxies as a function of redshift. This
results favours an 'equilibrium' model where the amount of turbulence of a disk
is defined by the balance between gas accretion and outflows.\medskip

\begin{figure}
\centering
\includegraphics[width=\linewidth]{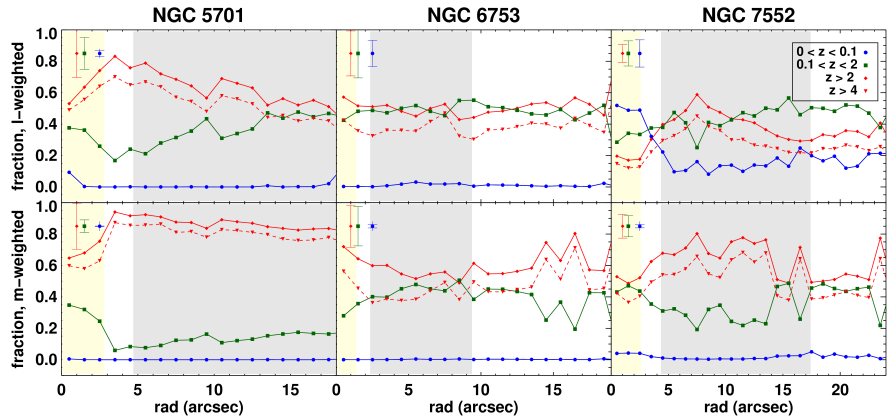}
\caption{Relative light (top row) and mass (bottom row) fractions of young,
intermediate and old stellar populations as a function of radius present in
three galactic bulges studied in \citet{2015MNRAS.446.2837S}. Uncertainties
in the analysis are indicated in the top left corner. Shaded regions mark the
regions where the average light and mass fractions of this study are computed.
More than 60\% of the stellar mass in those bulges was already in place beyond
$z\sim4$.}
\label{fig:9}
\end{figure}

The physical conditions between redshifts 1 and 4 appear to be particularly
favourable for the formation of bulges, and yet it appears that it cannot be the
only channel to build the (pseudo)bulges observed in the nearby Universe.
Mergers seem to be required too \citep[e.g.][]{2014arXiv1409.2622C}. To
complicate the issue further, the analysis of the star formation histories of
different types of bulges \citep[e.g.][]{2015MNRAS.446.2837S} suggest that at least
60\% of the stellar mass of those bulges formed at redshifts beyond 4 (see
Figure~\ref{fig:9}). All these results together indicate that bulge formation
most likely happens in a two stage process \citep[e.g.][]{2013ApJ...763...26O},
with an initial period of rapid build-up (with possible influence of mergers)
and a secondary phase (between redshifts 1 and 2) of high star formation
activity that would lead to the younger pseudobulge components we see today.

%-----------------------------------------------------------------------------
%-----------------------------------------------------------------------------

%=============================================================================
\section{Concluding Remarks \& Future Prospects}
\label{sec:5}

Lying at the centre and denser regions of galaxies, bulges are a keystone in our
understanding of galaxy formation and evolution. It is also their location,
shared with other components of galaxies what makes them so difficult to study.
In this review I have tried to provide an overview of the main kinematic
features observed in extragalactic bulges.\medskip

Identifying the formation scenario for bulges based solely on kinematic grounds
is a very difficult task. The orbits of the different structural components in
galaxies (e.g. bulges, disks, bars, spiral arms, nuclear disks rings, etc) are
not necessarily well separated in phase-space. The best example of this
complexity come from the observations of the Milky Way bulge. As nicely
illustrated in other contributions to this volume (e.g. Gonz\'alez \& Gadotti,
or S\'anchez-Bl\'azquez), the combined study of kinematics and stellar
populations provides one of the best ways to discern between different formation
scenarios.  While this coupling can be achieved relatively easy in the Milky Way
(because it is possible to measure the properties of individual stars) this is
no easy task in bulges of other galaxies where all we get is the integrated
light along the line-of-sight. Fortunately, with better data, models, and
numerical tools we are at the verge of being able to treat other galaxies in the
same way we study our own Galaxy. Studies of the coupling between kinematics and
stellar populations in external galaxies are now flourishing
\citep[e.g.][]{2008AN....329..980O}. Initially restricted to galaxies with known
distinct counter-rotating components, they are now exploring more regular
galaxies \citep[e.g.][]{2014MNRAS.441..333J}.\medskip 

As remarked many times throughout this review, this new step in the 3D
decomposition of galaxies can only be achieved with datasets that allow the
uniform exploration of galaxies in the two-dimensions they project in the sky.
The first generation of IFU surveys and instruments (e.g. SAURON, ATLAS3D,
DiskMass, SINFONI, VIMOS, PPaK) showed us the potential of these datasets to
reveal the intrinsic properties of galaxies. The currently ongoing IFU surveys
(e.g. CALIFA, SAMI, MaNGA, KMOS3D) will allow the exploitation of these new
techniques for very large, morphologically and mass unbiased samples of
galaxies. We should not forget though that we can still learn a lot of the
physical processes governing galaxies, and bulge formation and evolution in
particular, with unique instruments like MUSE. The Milky Way is a unique case,
as we will be able to probe the 3D nature of the Galaxy directly thanks to the Gaia space
mission.\looseness-2

%=============================================================================
\begin{acknowledgement}
J.~F-B would like thank D. Gadotti, E. Laurikainen and R.F. Peletier for their
invitation to take part in this volume and for their infinite patience waiting
for this review. J.~F-B acknowledges support from grant AYA2013-48226-C3-1-P
from the Spanish Ministry of Economy and Competitiveness (MINECO), as well as
from the FP7 Marie Curie Actions of the European Commission, via the Initial
Training Network DAGAL under REA grant agreement number 289313.
\end{acknowledgement}

%=============================================================================
\newpage
\bibliographystyle{mn2e}

\end{document}